# Biotransformation and biological impact of graphene and graphene oxide during simulated oral ingestion

Daniela Guarnieri[1], Paola Sànchez-Moreno[1], Antonio Esaú Del Rio Castillo[2], Francesco Bonaccorso[2], Francesca Gatto[1,3], Giuseppe Bardi[1], Cristina Martin[4,5], Ester Vázquez[4,5], Tiziano Catelani[6], Stefania Sabella[7*], and Pier Paolo Pompa[1*]

1) *Nanobiointeractions&Nanodiagnostics, Istituto Italiano di Tecnologia (IIT), Via Morego, 30 – 16163 Genova, Italy*

2) *Graphene Labs, Istituto Italiano di Tecnologia, Via Morego, 30 – 16136 Genova, Italy*

3) *Department of Engineering for Innovation, University of Salento, Lecce, Italy*

4) *Departamento de Química Orgánica, Facultad de Ciencias y Tecnologías Químicas, Universidad de Castilla-La Mancha, 13071 Ciudad Real, Spain,*

5) *Instituto Regional de Investigación Científica Aplicada (IRICA), Universidad de Castilla-La Mancha, 13071 Ciudad Real, Spain*

6) *Electron Microscopy Facility, Istituto Italiano di Tecnologia, Via Morego 30 – 16163 Genova, Italy*

7) *Drug Discovery and Development Department, Istituto Italiano di Tecnologia, Via Morego, 30 – 16136 Genova, Italy.*

**Abstract**

Graphene is an innovative nanomaterial, made of two-dimensional honeycomb-like carbon lattice, with potential in many different applications. Studying the behaviour of graphene-related materials (GRMs) in biological systems is, therefore, crucial to assess possible side effects. In this work is studied the biotransformation and biological impact of few layer pristine graphene (FLG) and graphene oxide ($GO_X$), following ingestion as exposure route. To mimic FLG and $GO_X$ ingestion, an *in vitro* digestion assay based on a standardized operating procedure (SOP) is applied. The assay simulates the human ingestion of GRMs during their dynamic passage through the different environments of gastro-intestinal (GI) tract (salivary, gastric, intestinal). Physical-chemical changes of GRMs during the digestion process are assessed through a detailed Raman spectroscopy characterization. Moreover, the effect of chronic exposure to digested GRMs on integrity and functionality of an *in vitro* model of intestinal barrier is also determined according to a second SOP. Our results showed that theFLG and $GO_X$ nanoflakes aggregates after experiencing the passage through the different environments of GI tract, with evident doping effects, due to the interaction of GRMs with digestive juice components and the strong pH variations. Interestingly, neither structural changes nor degradation of the nanomaterials are observed, suggesting that GRMs are biopersistent nanomaterials when administered by oral route. Chronic exposure to digested GRMs does not affect intestinal barrier integrity and is not associated to inflammation and cytotoxicity, though possible long-term adverse effects cannot be ruled out, due to the observed biodurability of the materials. Limited cellular internalization of digested GRMs is detected in the epithelial barrier, likely due to the formation of large GRM aggregates, with a prevalent localization in endo-lysosomes. These findings have important implications in the hazard identifications of GRMs upon ingestion, and pave the way to address their possible bio-applications.

**Keywords:** graphene, graphene oxide, digestive juices, biotransformation, intestinal barrier, inflammation, cytotoxicity

**Introduction**

In the latest years, the interest in graphene have grown continuously [1-2]. Graphene is a single layer of sp$^2$-hybridized carbon atoms tightly packed in a two-dimensional honeycomb lattice[1-2] (PNAS, Novoselov 2005). The group of graphene-related materials (GRMs) comprises single-layer graphene, few-layer graphene (FLG), graphene oxide (GO$_X$), reduced graphene oxide (rGO$_X$), graphene nanosheets, graphene nanoribbons, and graphene quantum dots [2-4]. GRMs have distinctive characteristics *e.g.* that make them interesting candidates for technological and biomedical applications, ranging from (opto)electronic to electrochemical devices, energy storage, cell imaging, drug delivery, and biosensors [3, 5-8]. Moreover, the use of graphene as nanofiller in food packaging has also been investigated because of its exceptional ability to limit oxygen permeation (some comparison with an state-of the-art material could aid to strength this idea) and light transmission in polymeric films (why is this important, also is necessary to put a number and/or comparison with a current used material) [9-11].

The integration of GRMs into consumer products makes crucial to assess their potential risk for humans, by defining their toxicological profiles and biological fate within the exposed organisms [12]. Exposure to GRMs can mainly occur via inhalation, ingestion, and/or skin contact. Amongst these, inhalation is considered as the most relevant way of entrance of GRMs in the human body and, hence, several *in vitro* and *in vivo* studies have recently focused their attention on this route of exposure [3, 13-14]. After inhalation, however, GRMs may also enter the digestive apparatus, through swallowing [15-17]. In addition, unintentional direct ingestion of GRMs could occur from contaminated waters or upon their release from food packaging. Despite such important route of entrance, few works on the fate and toxicological effects of GRMs upon oral exposure have been reported to date [18-21]. Moreover, *in vitro* models only partially mimic the real *in vivo* environment, *e.g.,* GRMs directly suspended in cell culture media without previous contact with GI juices [18] or pre-incubated with acidic buffers that only account for the low pH of the gastric compartment [19]. Therefore, *in vitro* data are poorly comparable to complex *in vivo* conditions, including strong pH shifts and variable concentrations of salts and enzymes during ingestion. Hence, a more reliable approach considering all steps occurring after oral ingestion is required for a realistic assessment of biotransformation of GRMs in the GI tract. In fact, as for other nanomaterials[22-25], the unique physical-chemical characteristics of GRMs may change depending on the surrounding conditions, such as temperature, pH, concentration, salts, etc, that, in turn, may modify the toxicological profile of GRMs on biological systems. Moreover, recent findings have shown that carbon nanomaterials, including $GO_X$ and oxidized carbon nanotubes (CNTs), may be degraded by oxidase enzymes, such as horseradish peroxidase (HRP) and human myeloperoxidase (hMPO) both *in vitro* and *in vivo* [26-28]. Notably, as also proposed in some recent categorization schemes for nanomaterials [29-30], a major concern for the potential harmfulness of GRMs is related to their biological transformation/persistence or release of toxic compounds through degradation.

Different digestion models have been applied to test changes in nanoparticle behaviour, mainly focusing on aggregation/agglomeration and cytotoxicity [31] [24, 32] [33]. In this work, we used a dynamic *in vitro* digestion assay, developed to mimic the human ingestion of nanoparticles and monitor their biotrasformations during the passage through the GI tract simulated environments (salivary, gastric, intestinal) [24]. The assay is part of a SOP [34] developed in the EU project NANoREG (A common European approach to the regulatory testing of nanomaterials. http://www.nanoreg.eu/). The application of SOPs has the scope to foster data reproducibility by lowering result variability, which often affects the benchmarking analysis among nanomaterials [35]. The assay had been validated using a range of reference nanoparticles (from JRC European repository list) for which a detailed physical-chemical characterization in pristine conditions is available (https://ec.europa.eu/jrc/en/scientific-tool/jrc-nanomaterials-repository). Here, we investigated biotransformation, biodurability in GI fluids, cell uptake, cytotoxicity, and inflammatory response of FLG and $GO_X$ upon the *in vitro* digestion. In particular, the first objective of this study is the understanding of the impact of each step of the *in vitro* digestion process on the physical-chemical properties of FLG and $GO_X$ flakes by Raman spectroscopy. The second objective is a thorough cytotoxicological investigation of digested GRMs on an *in vitro* model of intestinal barrier, a widely adopted test system by pharmacological industries and regulatory authorities [36]. A second NANoREG SOP method was used also for this analysis (De Angelis et al. "Standard Operating Procedure for evaluation of NPs impact on Caco2 cell barrier model" https://circabc.europa.eu/sd/a/fe1ea854-be73-4f58-a2ba-4bc2ce42ee62/SOP%20for%20the%20evaluation%20of%20NMs%20impact%20on%20Caco2%20cell%20barrier%20model%2029052015%20V9.pdf).

## Materials and Methods

*GRM synthesis and characterization*

Few layer pristine graphene (FLG) is prepared by ball-milling as described elsewhere [37]. Monolayer/few layer graphene oxide ($GO_X$) is provided by Grupo Antolin Ingeniería (Burgos, Spain). GRM morphology and lateral size are analyzed by Jeol JEM 1011 transmission electron microscope (TEM) (Jeol, Japan). Thermogravimetric analysis (TGA) is carried out using a TGA Q50 (TA Instruments) at 10 $°C$ $min^{-1}$ under nitrogen flow, from 100 °C to 800 °C. Measurements of zeta potential of FLG and $GO_X$ are carried out with a Zetasizer Nano-ZS (Malvern Instruments, Worcestershire, UK) at 25 °C in low ionic strength buffers at different pH. All measurements ae performed in triplicate for each sample.

*In vitro digestion assay*

The digestion of GRMs is carried out by an *in vitro* digestion assay that simulates the human ingestion of nanomaterials. The assay is based on a dynamic model developed by Bove et al.[24] that is also available as a SOP [34] from the European project Nanoreg (NANoREG – A common European approach to the regulatory testing of nanomaterials. http://www.nanoreg.eu/). Briefly, the assay employs artificial juices simulating the human digestive compartments (mouth, stomach and small intestine), which are dynamically added into the Eppendorf tube under stirring conditions. The assay is slightly modified. The digestive juices are prepared in sterile conditions by combining salt solutions, organic compounds and proteins to obtain the final concentrations in a total reaction volume of 10 ml as reported in Table S1. The final pH of each single juice is 6.8 ± 0.1 for saliva, 1.3 ± 0.1 for stomach, 8.1 ± 0.1 for duodenal and 8.2 ± 0.1 for bile. The juices are pre-heated to 37 °C for at least two hours before starting the experiments. All chemicals were purchased by Sigma Aldrich.

The assay is conducted following the reported procedure: 20 µl of the FLG and $GO_X$ dispersions (0.09 and 0.45 mg ml$^{-1}$, respectively) are added into a 1.5 mL Eppendorf tube. Afterward, the digestive juices are added in a temporal sequence that simulated the transit of food bolus along the gastrointestinal apparatus [38]. To reproduce the mouth compartment, 60 µl of salivary juice at pH 6.8 are mixed with 20 µl of the GRM dispersion and shacked at 37 °C for 5 minutes. After incubation, 1 µl of the mouth sample was collected and dried on a silicon wafer for Raman spectroscopy analysis (described below). The remaining sample is used to continue the transit into the stomach. To this aim, 120 µl of gastric juice are added to the mouth samples, the pH is adjusted to 2.5 ± 0.5 with 1 M NaOH and the samples are incubated for further 120 minutes at 37 °C under shaking. At the end of gastric digestion, 1 µl of the stomach samples is processed for the Raman analysis. The remaining samples are employed to simulate digestion in the small intestine, adding to it 120 µl of duodenal fluid, 60 µl of bile salts and 20 µl of 84.7 g l$^{-1}$ sodium bicarbonate solution and adjusting the pH at 6.5 ± 0.5 with 3.7% HCl. The shaking is stopped after further 120 minutes of incubation and 1 µl of the samples was dried on a silicon wafer for Raman characterization. For biological experiments with Caco-2 cell layers, the digestion process is carried out in sterile conditions and the digested GRMs are diluted 1:5 in cell culture medium before incubation with the epithelia.

*Raman spectroscopy*

1 µL of dispersion for each sample was drop cast on a Si wafer (LDB Technologies Ltd.) coated with 300 nm of thermally grown $SiO_2$. The Raman spectra are measured using a Renishaw confocal microscope (514.5 nm laser excitation wavelength with an incident power of ~1 mW on the sample), with a 50× objective and a grating of 2400 l/mm. The deposited samples are mapped in rectangular areas of ~100 µm × 100 µm. The offset between points in the mapping is set to 5 µm. For statistical analysis, 50 spectra of GRM are selected on each mapping-sample. The FLG and $GO_X$ peaks were fitted with Lorentzian functions, all the spectra are normalized to the integral intensity of the G band.

*Intestinal layer formation and chronic treatment with digested GRMs*

Human colon epithelial (Caco-2) cells (gently provided by Dr. Isabella De Angelis, Istituto Superiore di Sanità (ISS), Rome, Italy) are cultured in Dulbecco's Modified Eagles Medium (DMEM, Sigma-Aldrich) supplemented with 10% (v/v) fetal bovine serum (FBS, Sigma-Aldrich), 1% non-essential aminoacids (Invitrogen), 100 U ml$^{-1}$ penicillin and 100 mgml$^{-1}$ streptomycin (Sigma-Aldrich). Cells were maintained in incubator at 37 °C under a humidified controlled atmosphere and 5% $CO_2$. To obtain intestinal epithelia, cells are seeded in 12-well plates onto porous Millicell hanging cell culture inserts (Merck Millipore) (*d*, 12mm; *A*, 1.1cm$^2$; pore size 0.1 µm) made of polyethylene terephtalate (PET) in 0.5 mL of medium at a seeding density of 1.7 x 10$^5$ cells/insert in the apical side. 1.5 mL of medium are poured in the basolateral compartment. Cells are grown for three weeks, and culture medium is changed every two days, to allow the formation of tight junctions and microvilli according to the NANoREG SOP "Standard Operating Procedure for evaluation of NPs impact on Caco2 cell barrier model". Before starting each experiment, Trans-Epithelial Electrical Resistance (TEER) is measured to verify the correct formation of confluent intestinal layers. The cell inserts are then incubated with digested FLG and $GO_X$ diluted 1:5 in cell culture medium at the final concentrations of 1 and 5 µg ml$^{-1}$, respectively, for 2 hours every 2 days up to 9 days in order to mimic a chronic intestinal exposure. As a control, some cell inserts are incubated with digestive juices without GRMs at the same conditions used for digested materials to verify their possible effect on cell layer integrity and functionality.

*Trans-Epithelial Electrical Resistance (TEER) measurements*

Before and after 1, 5 and 9 days of incubation with digested GRMs, integrity of differentiated Caco-2 cell epithelia are evaluated by TransEpithelial Electrical Resistance (TEER) using a chop-stick electrodes device (Millicell-ERS voltmeter, Millipore). TEER values are expressed as Ohms ($\Omega$) x $cm^2$ and are calculated according to the following equation(REFERENCE):

$$TEER = [\Omega \text{ cell monolayer} - \Omega \text{ filter (cell-free)}] \times \text{filter area } (1.12 \text{ cm}^2)$$

Inserts are considered suitable for experiments if TEER value is >150 $\Omega$ x $cm^2$.

*Lucifer yellow (LY) assay*

At the end of experiments, the impact of digested GRMs on epithelium integrity is evaluated by lucifer yellow (LY, Sigma) assay to determine any difference in this paracellular marker ability to cross the monolayer between GRM-treated inserts and untreated inserts. After 1, 5 and 9 days of incubation with digested GRMs, apical (Ap) and basolateral (Bl) media are collected and cell layers are washed twice with HBSS. Ap compartment was filled with 0.5 ml of 0.4 mg $ml^{-1}$ LY solution in HBSS and Bl compartment with 1.5 ml HBSS. Cells are then incubated for 2 h at 37 °C. After incubation, 100 µl of the Bl HBSS of each insert (including free cell inserts) are collected and added into a black 96-well plate. LY content is measured by fluorometric detection (ex. 428 nm, em. 536 nm). The percentage of LY passage in Bl side after treatment is compared to the percentage of LY passage in the negative control.

*Indirect immunofluorescence and confocal microscopy*

After chronic incubation with digested GRMs, cell layers are fixed with 4% paraformaldehyde for 20 min at room temperature, permeabilized with 0.01% Triton X100 for 5 min and blocked with blocking buffer solution (0.5% bovine serum albumin in PBS) for 20 min. Cells are then stained with 0.1 nM Alexa Fluor™ 594 Phalloidin for 30 min and Hoechst 33342 (Thermo Fisher Scientific) at a concentration of 5 µg $ml^{-1}$ for 5 minutes, to localize actin microfilaments and cell nuclei, respectively . Lysosomes are localized by using anti-LAMP1 primary antibodies (Abcam, ab24170) and Alexa546 anti-rabbit secondary antibodies (Thermo Fisher Scientific). Cells are incubated with primary antibodies in blocking buffer solution (1:200) for 1 hour and after several washes with PBS, cells are incubated with the secondary antibodies (1:500) for 45 min. Confocal microscopy images are acquired by a confocal microscope (Leica TCS-SP5) with an oil-immersion 63× objective, 405, 488 and 561 nm excitation laser wavelengths and a resolution 1024 × 1024 pixels. Z-sectioning images are acquired with a z-slice thickness of about 0.7 μm.

*Transmission electron microscopy (TEM)*

To observe the formation of microvilli and tight junctions as well as the intracellular localization of digested GRMs, the Caco-2 barriers are fixed for 2 h in 1.5% glutaraldehyde in 0.1 M Sodium Cacodylate buffer (pH 7.4), post fixed in 1% osmium tetroxide in the same buffer and stained overnight with 1% uranyl acetate aqueous solution. The barriers are then dehydrated in a graded ethanol series, infiltrated with propylene oxide and embedded in epoxy resin (Epon 812, TAAB). Semi-thin and thin sections of the embedded cell monolayer are cut with an ultramicrotome (UC6, Leica) equipped with a diamond knife (Diatome). Images are collected with a Jeol JEM 1011 (Jeol, Japan) electron microscope, operating at an acceleration voltage of 100 kV, and recorded with a 11 Mp fiber optical charge-coupled device (CCD) camera (Gatan Orius SC-1000).

*Cell viability assay*

Cell viability is evaluated by measuring the cell metabolic activity using CellTiter 96® AQueous One Solution Cell Proliferation Assay ((3-(4,5-dimethylthiazol-2-yl)-5-(3-carboxymethoxyphenyl)-2-(4-sulfophenyl)-2H-tetrazolium, inner salt) MTS assay, Promega). Briefly, $5\times10^3$ cells suspended in 100 μl of cell culture medium are seeded in each well of a 96-well plate. Following 3 days of culture, cells are treated with 1 μg ml$^{-1}$ of digested FLG and 5 μg ml$^{-1}$ of digested GOX diluted 1:5 in cell culture medium for 2 h every day, up to 4 days of culture. 0.05 mg ml$^{-1}$ of benzalkonium chloride (BC, Sigma) in cell culture medium was used as a chemical positive control. Cell treatments with non-digested FLG and GOX at the same concentrations of digested GRMs and with digestive juices alone are used as further controls. After at 24, 48 and 96 h of treatment, cell culture medium of each well is replaced with 100 µl of fresh medium supplemented with 20 µl of MTS reagent and incubated for two hours at 37 °C. The optical absorbance is measured at 490 nm using a microplate reader (Synergy HT, BioTek) and the raw data are normalized to non-treated cells (considered 100%) to calculate cell viability percentage. Experiments are performed in triplicate and data are reported as mean ± standard deviation.

*LDH assay*

After cell interaction with digested GRMs as described above, to assess the impact on cell membrane, the lactate dehydrogenase (LDH) leakage assay is performed onto 96-well microplates by using the CytoTox-ONE Homogeneous Membrane Integrity Assay reagent (Promega), following the manufacturer's instructions. LDH released in the extracellular environment is measured with a 10 min coupled enzymatic assay that results in the conversion of resazurin into fluorescent resorufin (560 Ex/590 Em) by using a plate reader (Synergy HT, BioTek). As a negative control, the same assay onto untreated cells is performed. To determine the maximum LDH release (positive control), 2 µl of Lysis Solution are added to positive control wells 10 minutes before the assay was performed. Results are normalized with respect to positive control (expressed as 100 %).

*Cytokine and chemokine release*

Inflammatory cytokine release (panel: IL-8, MCP-1, IL-1β, IL-6, INFγ, TNFα, MIP1β, RANTES) in the apical and basolateral media from the Caco-2 cell layers after 1, 5 and 9 days of chronic treatment with digested GRMs are assessed with a Bio-Plex 1 MAGPIX TM Multiplex Reader (Bio-Rad) according to the manufacturer's procedure. The cells were stimulated with 100 ng ml$^{-1}$ lipopolysaccharide (LPS, from Escherichia coli 0111-B4, Sigma–Aldrich, cat. no. L4391) as positive control.

**Results and Discussion**

*Biotransformation of GRMs during the digestion process*

A schematic representation of the *in vitro* digestion assay adapted for GRM ingestion is reported in Figure 1 and Table S1. This proposed model ~~mimics the gastro-intestinal passage and~~ simulates the oral, gastric, and small intestine conditions. To mimic the gastro intestinal process, synthetic digestive juices are used, and the pH changes, transit times, relevant enzymes, and protein compositions during the digestion process are taken into account, as described in previous reports [24,38]. In this study, FLG and GO$_X$ flakes are used as model GRMs. FLG flakes are obtained by exfoliation of graphite through interaction with melamine by ball-milling treatment [37]. After exfoliation, melamine is removed by filtration with hot water (write temperature) to obtain stable dispersions of FLG. The final FLG concentration as estimated to be 0.09 mg ml$^{-1}$ with melamine traces (0.09 ppm). GO$_X$ flakes are obtained through oxidation of carbon fibers (GANF Helical-Ribbon Carbon Nanofibers, GANF). To remove the presence of acids, the initial GO$_X$ suspensions (concentration < 1 mgml$^{-1}$) are rinsed with MilliQ water by centrifugation, at 4000 rpm (write the Centrifuge or rotor model or "g" force) for 30 min, until a pH value of ~5 is reached.

The aggregation state and the surface charge of nanomaterials influence their biological behaviour. Therefore, the as-prepared GRMs are characterized by transmission electron microscopy (TEM), thermogravimetric analysis (TGA), and zeta potential. TEM analysis reveal a broad lateral size distribution of both GRMs (Figure 2 c and d), in line with previously reported results [39], the mean flake length is similar for both materials c.a. 400 nm. Representative flakes, with many visible wrinkles, of FLG and GO$_X$, are shown in Figure 2 a and b, respectively. Thermogravimetric analysis reveals a weight loss of 8% for FLG at 600 °C, indicating a low quantity of oxygen groups generated by the exfoliation process, while a weight loss of 46% in the case of GO$_X$ at the same temperature (Figure 2 e and f). Moreover, zeta potential indicated a negative charge around -20 mV for FLG and -35 mV for GO$_X$ in MilliQ water (Figure 2 g and h).

According to their route of entrance in biological systems, *e.g.* XX,YYY, zZZ, nanomaterials experience different environments that affect their original chemical/physical properties.(REF) For instance, it has been reported that, during digestion, silver nanoparticles are fully dissolved to silver ions that, in turn, interact with the components of digestive juices, forming secondary silver-organic complexes [24]. Moreover, metal containing nanoparticles, once internalized by cells according to endocytosis pathways, can be degraded in the lysosomal compartment, due to the low pH and degradative environment [22-23, 40]. It has also been recently reported that FLG and $GO_X$ can be degraded by peroxidases (hMPO and HRP) *in vitro* [26-28].

Starting from these observations, Raman spectroscopy analysis is carried out on GRMs incubated at specific time intervals in the different digestive juices, to assess their biotransformation/biodegradation in conditions that mimic the digestion process. Raman spectroscopy has demonstrated to be a powerful tool for the characterization of nanomaterials, *e.g.*, in terms of doping [41-43], functionalization [44], oxidation [45-46]. In particular, for GRMs, Raman spectroscopy has been widely used to identify the number of layers, defects type and doping, disorders on the crystalline structure, chemical modifications, just to cite a few [47-50]. Thus, Raman analysis provide important information about the possible chemical and physical changes in the graphene flakes as they go through the simulated digestive tract. The Raman spectrum of graphene is composed by the G peak, positioned at $\sim$1580 cm$^{-1}$, the D peak, positioned at $\sim$1350 cm$^{-1}$, and the 2D peak, centered at $\sim$2680 cm$^{-1}$ in case of a single layer graphene. A detailed physical description of the main Raman modes of graphene is reported in the Supporting information. **Figure 3** shows representative Raman spectra of the FLG, $GO_X$, and the spectra of the different digestive juices: saliva, stomach and intestine. We can observe the characteristics D, G, D' and 2D bands for FLG spectra, as well as the typical broad D and G bands, characteristics in $GO_X$. The Raman spectra of the physiological juices are characterized by two main bands cantered at 1450 and 2900 cm$^{-1}$.

Doping (*e.g.*, electrostatic, metallic contacts, chemical, heteroatoms or optical) strongly affects the G, D, and 2D band lineshapes/positions[45, 51-53]. In particular, the position and full width at half maximum (FWHM) of the G band (pos(G) and FWHM(G), respectively), change due to electron-phonon coupling and the Kohn anomaly at the Γ point[42, 50]. The frequency of the G band has the lowest value when the Fermi level is at the Dirac point, and the pos(G) stiffens as the doping concentration increases[43]. Contrary, the FWHM(G) decreases as the doping concentration increases[51-52]. While the G band obeys to the doping concentration, the 2D band is sensitive to the doping type, this means that pos(2D) increases if the doping is p-type and decreases for n-type doping[51]. The 2D band intensity, I(2D), is also affected by the doping concentration, but it is insensitive to the doping type, decreasing with the increase of doping[43, 45]. In the same way, as the 2D band is doping-sensitive, the first order D band is doping-sensitive as well[54]. The I(D)/I(G) reaches the minimum value when the Fermi level (which can be tuned according to the amount and type of dopants attached to the graphene) approaches the Dirac point. I(D)/I(G) reaches the maximum value when the Fermi energy is at half of the excitation energy[55], *i.e.,* the Pos(D) up-shifts with p-type doping and down shifts with n-type doping[55]. Moreover, also the FWHM increases with n-type doping and decreases for the p-type one[55]. Structural defects, aside from being the responsible for the D band activation[49, 56], also modify significantly the pos(G) and FWHM(G)[46, 57], *i.e.,* for a graphene flake with structural defects the FWHM(G) is larger than 24 cm$^{-1}$ [46, 57], and the pos(G) can be shifted up to 5 cm$^{-1}$ [46, 57]. Additionally, the FWHM of all the graphene Raman peaks, *i.e.*, D, D', G and 2D, increases with the presence of defects, while the intensities of the D and D' increases and the intensity of the 2D decreases in the presence of defects[54].

Combined to the fact that the electronic and optical properties of graphene change due to defects and doping, the Raman spectra also change with the number of stacked graphene layers[47, 58-59], i.e., the 2D band changes in lineshape and position for an increasing number of layers[47, 58]. These variations in the Raman spectra make the data interpretation challenging for a heterogeneous sample, as it is the case of the mechano-chemical exfoliated graphene or GO$_X$ flakes. As a consequence, an extensive statistical analysis is a compulsory step, in order to understand the physical/chemical changes of graphene/GO$_X$ dispersed in biological media.

The most significant results of the Raman statistical analysis on FLG and GO$_X$ are shown in **Figures 4 and 5**. In Figure 4a, the Pos(G) of the FLG (~1582 cm$^{-1}$) is maintained in saliva and stomach samples. The Pos(G) in intestine down-shifted to ~1580 cm$^{-1}$, due to the doping influence from the constituents of the digestive juices. The doping hypothesis is corroborated observing the FWHM(G) of the samples (**Figure 4b**), which stiffen from 22 to 18 cm$^{-1}$, when the FLG passed from saliva to stomach/intestine. In the normalized intensity of the D band (I(D)/I(G)) (**Figure 4c**), the maximum population is found at 0.3, which is constant both saliva and intestine. Then, the I(D)/I(G) distribution grows in stomach due to the contribution of small flakes that aggregate in the presence of low pH, as also demonstrated by Zeta-potential titration measurements (**Figure S1, right panel**). In particular, at highly acidic pH (1.2), Zeta-potential values of FLG are close to 0 mV, thus falling in the range of instability of colloidal suspensions [60]. The FWHM(D) (**Figure 4d**) ranged from 32 to 60 cm$^{-1}$ for the FLG, indicating that the sample is composed by a broad flake size distribution, since FWHM(D) increases inversely with the crystallite size[61-62]. Notably, the FWHM(D) for the FLG dispersed in saliva and intestine have the same distribution as the one of the pristine sample, while in stomach juice the distribution decrease at 30 to 50 cm$^{-1}$, due to the protonation of the FLG places (p-type doping)[45, 51-53]. The formation of structural defects, *i.e.*, creation of sp$^3$ hybridized carbon atoms, on the FLG is discarded by the fact that the I(D)/I(G) distribution in intestine peaked at 0.4 and spanned from 0 to 1.1, which is the same distribution of the starting FLG sample; a similar behaviour is observed for the FWHM(D). The 2D sub-components analysis (Figure S2) showed no changes in pos(2D$_1$) and pos(2D$_2$). The intensities of the sub-peak 2D$_1$, I(2D$_1$)/I(G) increase from 0.3 to 0.6 when the pH value changed from neutral to acidic (stomach), which led us to conclude that the 2D$_1$ intensity is affected by the doping. The intensities of the 2D$_2$, I(2D$_2$)/I(G), remained unchanged for all the samples, despite the changes in doping.

The main GO$_X$ Raman peaks are the G and D bands. Statistical analysis of the GO$_X$ Raman data made possible to observe the Pos(G) (~1580 cm$^{-1}$) (**Figure 5a**) and FWHM(G) (**Figure 5b**) remained unchanged for all the samples, except for the FWHM(G) in intestine, in which the distribution became narrow, shifting the maximum population distribution from 100 to 70 cm$^{-1}$. The stiffening of the G band is ascribed to the doping contribution (the charge impurities, *i.e.*, dissolved salts and H$^+$)[51].

Observing the I(D)/I(G) and FWHM(D) for $GO_X$ in intestine and stomach (**Figure 5c** and **d**, respectively), an intensity reduction and stiffening in the band are clear. Here, the aggregation of the flakes and the doping, due to the pH changes as well as the adsorption of juice components like proteins are the responsible of this behaviour [63] (**Figure S1 right panel**).

In summary, Raman analyses indicated that the digestive process does not induce structural defects in FLG, as no significant changes are observed in the I(D)/I(G) and FWHM(D) in the final step (intestinal juice). In addition, the changes of the peaks position (D, G and 2D), intensity, and linewidth are mainly due to changes in doping, likely because of pH changes. A partial role, however, can be also ascribed to the other components of the digestive juices, such as proteins and organic molecules that can interact/adsorb to FLG and $GO_X$ flakes, forming the so-called protein corona and affecting their stability and surface chemistry [25, 64]. For the $GO_X$, neither defects are induced nor reduction of GOX are observed. Raman observations thus indicate that, when GRMs experience the digestive environment, including strong pH variations, degradative enzymes, ions and other organic molecules, they do not undergo significant biodegradation processes, unlike other nanomaterials [24, 65], suggesting their possible biodurability in the GI tract with unpredictable long-term effects. Moreover, the evident doping can change the nature/chemistry of starting materials thus influencing the interaction with biological systems.

*Assessment of intestinal epithelium integrity upon chronic exposure to digested GRMs*

The intestinal barrier is one of the most important biological barriers within the human body. It attends to several functions, such as nutrient uptake, protection against pathogens, and preservation of intestinal microbiome [66]. The impairment of intestinal homeostasis can lead to uncontrolled entrance of pathogens and food antigens as well as in dysregulated nutrient supply, which in turn can compromise the health of the entire organism.(REF) Therefore, as a final step of the digestion process, we test the biological impact of the digested GRMs on intestinal barrier according to a second SOP (https://circabc.europa.eu/sd/a/fe1ea854-be73-4f58-a2ba-

4bc2ce42ee62/SOP%20for%20the%20evaluation%20of%20NMs%20impact%20on%20Caco2%20cell%20barrier%20model%2029052015%20V9.pdf). To achieve a reliable *in vitro* model of intestinal barrier, human intestinal epithelial (Caco-2) cells are grown for three weeks on porous inserts (Figure 6a). Confocal z-sectioning and TEM images show the formation of a confluent cell layer with the typical structures of intestinal epithelium,(REFS) such as cell-cell junctions and microvilli (Figure 6b and c). Intestinal barriers are then exposed (up to 9 days) to digested FLG and $GO_X$, at the final concentrations of 1 and 5 μg ml$^{-1}$, respectively. To choose the dose of GRMs, we referred to the human dietary uptake of other nanomaterials.(REF) In fact, for silver nanoparticles, concentrations ranging from 1 to 100 μg ml$^{-1}$ were considered to be a realistic dose range *in vitro* [65, 67-68]. Additionally, the used FLG and $GO_X$ concentrations are selected in order to have the maximum possible concentration, allowing good dispersion/stability of the nanomaterials (in particular, the concentration of pristine FLG is lower than $GO_X$, due to its worse dispersibility in water, Z-potential of 28 mV at pH 7 for FLG and 40 mV for $GO_X$ see Figure S1).

The integrity of the intestinal layer upon chronic incubation with digested GRMs is assessed by measuring Trans-Epithelial Electrical Resistance (TEER) and passage to the Bl compartment of LY, a marker of paracellular transport (Figure 6d and e). Epithelial layers treated with digested GRMs do not show detectable differences in TEER and LY with respect to non-treated controls (Figure 6d and e). These results indicated that digested GRMs are well tolerated by the intestinal barrier and does not induce its disruption/perturbation upon chronic exposure. Similar results were also reported by Bohmert et al. after treatment of Caco-2 cell layers with digested silver nanoparticles [65]. In that case, no variations in impedance measurements were observed up to 24 h prolonged incubation.

*Cellular uptake and intracellular localization of digested GRMs*

Since digested GRMs do not compromise the integrity of the intestinal barrier, their capability to be internalized by Caco-2 cell layers was investigated by confocal microscopy after 9 days of chronic incubation. As shown in Figure 7a and d, very few spots of both FLG and $GO_X$ are observed in the intestinal layers. Confocal z-sectioning of intestinal epithelia ~~stained with phalloidin and Hoechst~~ confirmed, from a morphological point of view, that the treatment with digested FLG and GO$_X$ do not affect cell layer integrity and demonstrated the intracellular localization of GRM spots. Under experimental conditions, we observe a limited internalization of GRMs after digestion process, which is due to the GRM aggregation when are in contact with the components of digestive juices, as suggested by Raman analysis. In fact, we observe large GRM aggregates associated to the intestinal barriers and, in some cases, with a preferential accumulation on the cell membrane, in particular along cell boundaries in intestinal barriers (Figure S3). The size of these aggregates is variable and around some microns (Figure S3). Another important factor that may reduce the cellular uptake of GRMs is the differentiation status of Caco-2 cells as enterocytes. In fact, it is reported that the undifferentiated Caco-2 cells are more prone than differentiated ones to internalize nanomaterials, such as GO$_X$, quantum dots, TiO$_2$ and SiO$_2$ nanoparticles [21, 69-71]. In particular, Kucki et al. recently demonstrated that the very dense brush border (made of microvilli) of differentiated intestinal epithelium led to low adhesion of GO$_X$ sheets and steric hindrance for material uptake [21]. However, a clear understanding of GRM internalization into the cells has not yet been achieved, due to the poor internalization and the difficulty to have comparable *in vitro* models [21, 72-75]. In TEM images is further confirmed the internalization in Caco-2 cell barriers and suggests a vesicular confinement of internalize digested FLG (Figure 7 b and c) and GO$_X$ (Figure 7 e and f). In general, most nanomaterials use endocytosis mechanisms to penetrate the cell membrane and, consequently, they preferentially accumulate into endo-lysosomes [23, 39, 69, 76]. Therefore, to elucidate the mechanisms of cellular uptake of digested FLG and GO$_X$, the lysosomal localization of internalized GRMs within the intestinal barriers is analysed. Immunofluorescence results indicate the co-localization of digested GRMs with the lysosomal marker LAMP1, thus suggesting that endocytosis contribute to the internalization of FLG and GO$_X$ flakes (Figure S4).

Because of the degradative environment of the lysosomal compartment where digested FLG and GO$_X$ are localized after cellular uptake, additional studies to assess biocompatibility of GRMs on undifferentiated/proliferating Caco-2 cells were carried out. Although only differentiated Caco-2 cells represent the epithelial cell layer of the small intestines, immature intestinal cells are also present in the small intestines, due to cell renewal, and they are proved to be more sensitive to external disturbances, *e.g.*, when exposed to silver nanoparticles after *in vitro* digestion [65]. Therefore, viability and cell membrane integrity of undifferentiated Caco-2 cells is investigated by 3-(4,5-dimethylthiazol-2-yl)-5-(3-carboxymethoxyphenyl)-2-(4-sulfophenyl)-2H-tetrazolium (MTS) and lactate dehydrogenase (LDH) assays, respectively. To this aim, undifferentiated Caco-2 cells are exposed to digested FLG and $GO_X$ for 2 hours every day, up to 4 days. Confocal microscopy images demonstrate the internalization of digested GRMs in Caco-2 cells (Figure S5a). As expected, the uptake of digested GRMs by undifferentiated Caco-2 cells was significantly higher than intestinal barriers, because, contrary to the latter, undifferentiated Caco-2 cells lacked the typical structures, such as tight junctions and microvilli, which hinder nanomaterial internalization [21, 69-70]. Despite such higher uptake, both cell viability and cell membrane integrity are not affected by the treatment (Figure S5 b and c). Moreover, treatments with non-digested GRMs, used as controls, further indicated neither significant decrease in cell viability, nor damage of cell membrane (Figure S5 b and c), in line with previously reported data on the same cell type [18-19]. Therefore, lack of cytotoxicity in these cells after exposure to digested GRMs may likely suggest no toxicity on intestinal barrier.

*Inflammatory response of intestinal epithelium to digested GRMs*

*In vivo* studies on laboratory animals provided some indications that inflammation may be involved in the toxicity of GRMs upon inhalation, and that the extent of inflammatory response could depend on the physical-chemical characteristics (i.e. lateral size, oxidation) of GRMs [13-14]. Therefore, the release of inflammatory cytokines in the apical and basolateral media is measured to evaluate the possible triggering of inflammation by the intestinal layers upon chronic exposure to the digested GRMs. Caco-2 cell layers usually show a significant increase of IL-8 and MCP-1 levels when stimulated with inflammatory agents [77-79]. In our experiments, the levels of IL-8 and MCP-1 are comparable to untreated cell layers used as negative control (Figure 8). On the contrary, positive stimulation with LPS increase the release of the two cytokines (Figure 8). Hence, at variance with other nanomaterials [80], digested GRMs do not induce any significant pro-inflammatory effect on the intestinal epithelium *in vitro*. A possible explanation for this effect could be related to the larger dimensions of GRMs after digestion, due to aggregation. In particular, GRM aggregates could have different transport rates and be retained outside the cell layer, unlike smaller nanomaterials or ions that may cross the intestinal layer and reach the cells more easily, inducing stronger inflammatory effects. This is consistent with previous findings using Ag nanoparticles, where only smaller 20 nm particles up-regulated the IL-8 expression in Caco-2 cell layers, while bigger 100 nm Ag nanoparticles did not [77-78]. Thus, the immobilization of the large GRM aggregates, due to the size-exclusion by the intestinal barrier, could result in a reduced cellular uptake and low cytokine release.

## Conclusions

In this work, we investigated the biotransformation and biological impact of few layers graphene and graphene oxide flakes upon ingestion, by using the NANoREG standard methods simulating *in vitro* digestion. Our results highlighted the influence of digestive juices in modulating few layers graphene and graphene oxide physical-chemical properties. In particular, the interaction of both materials with ions and other molecular components present in digestive juices resulted in evident doping effects and no structural changes. This interaction influenced the aggregation state of few layers graphene and graphene oxidewith important consequences in bioaccessibility of these materials to the intestinal layer. In fact, digested GRMs were well tolerated by the intestinal barrier up to 9 days of exposure, not inducing detectable damage, even though large GRM aggregates were associated to its apical side. The immobilization of the GRM aggregates, due to the size-exclusion by the typical brush border of the intestinal barrier, resulted in: i) reduced cellular internalization, ii) no short-term cytotoxicity and iii) low cytokine release. However, because of the observed GRMs biodurability, regardless of the complex and harsh environments they experienced during the digestion simulation and their partial cellular uptake, additional investigations on their long-term fate are necessary in future studies to further assess their biocompatibility profile.

**Acknowledgements**

The authors acknowledge Dr. P. Bove and Dr. C. Carnovale for their help during experiments and also the Antolin Group for supporting the commercial material. This project has received funding from the European Union's Horizon 2020 research and innovation program H2020-Adhoc-2014-20 under grant agreement No. 696656 – GrapheneCore1 and from the Spanish Ministry of Economy and Competitiveness MINECO (project CTQ2014-53600-R and FEDER fonds).

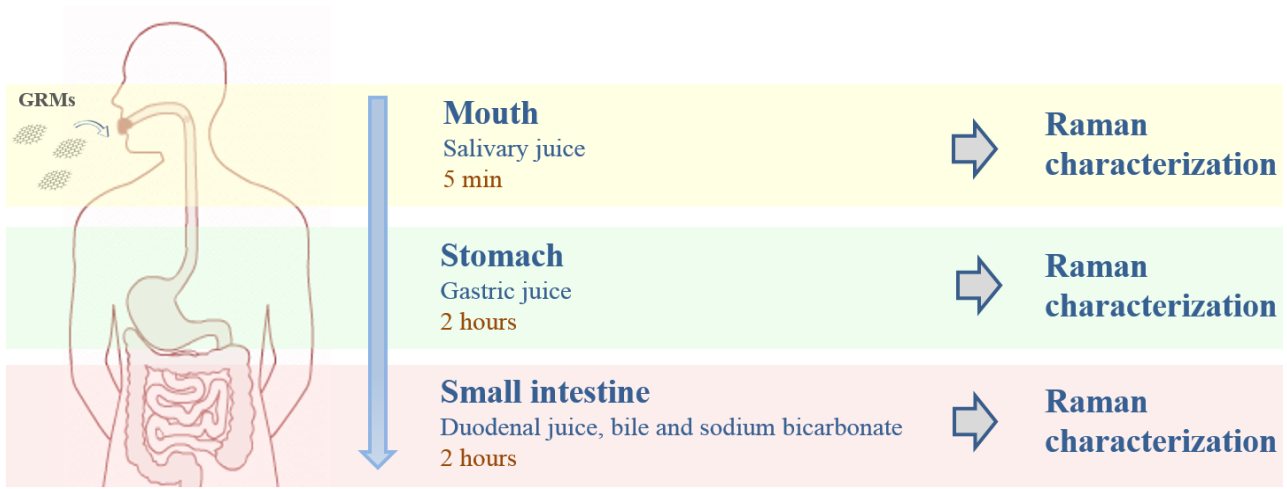

**Figure 1. Schematic representation of the *in vitro* digestion assay of GRMs.** Along the passage through the different digestive compartments (mouth, stomach and small intestine), chemical-physical changes of FLG and $GO_X$ were monitored by Raman spectroscopy. Synthetic digestive juices are used and all relevant parameters during digestion process such as temperature, pH changes, transit times, relevant enzymes, and protein compositions are considered.

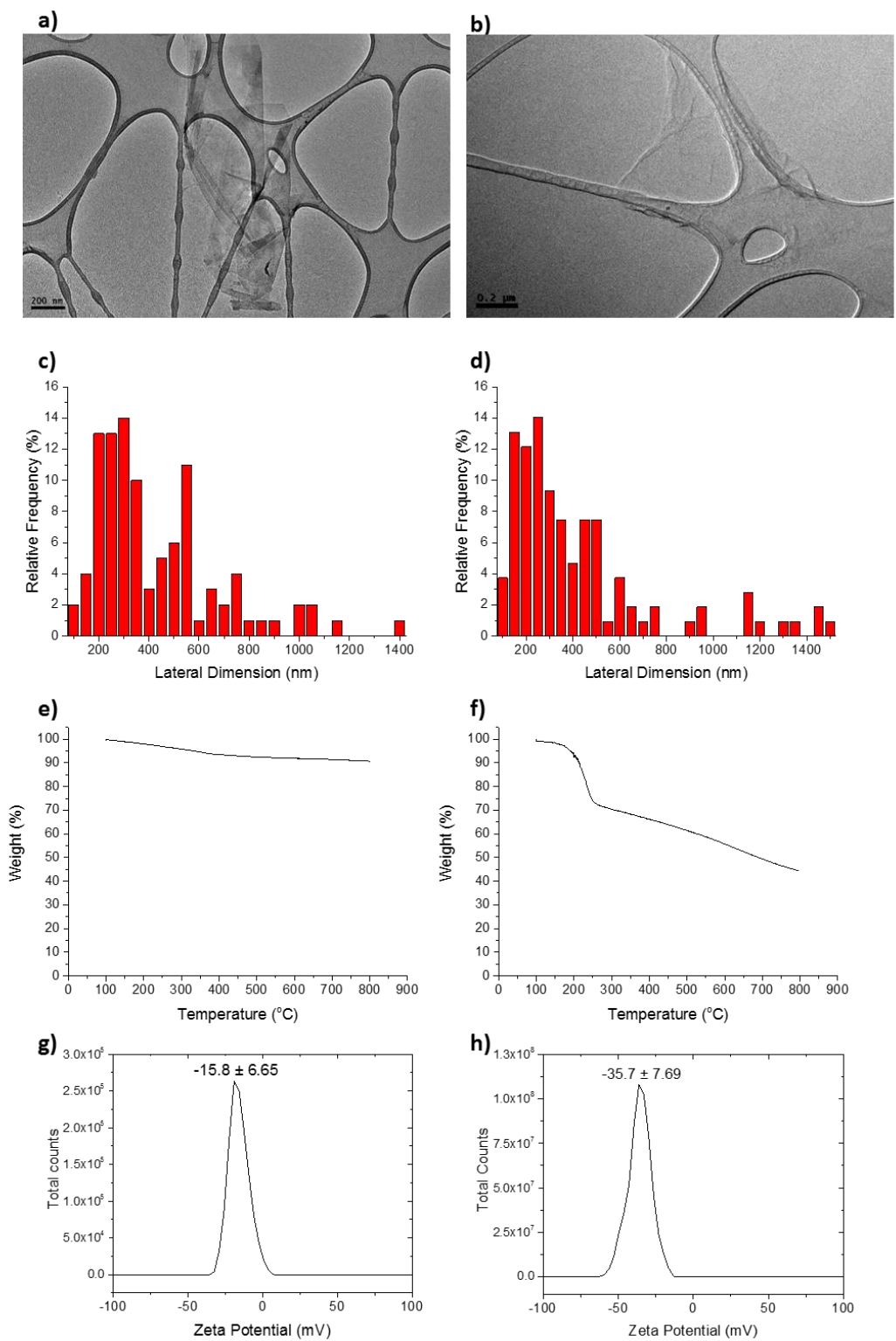

**Figure 2. Graphene and graphene oxide characterization.** Representative transmission electron microscopy images of FLG (a) and $GO_X$ (b) nanosheets. Lateral dimension distribution of FLG (c) and $GO_X$ (d) flakes measured by TEM image analysis. Thermogravimetric analysis of FLG (e) and $GO_X$ (f) Zeta potential of FLG (g) and $GO_X$ (H) measured at 25 °C and dispersed in MilliQ water.

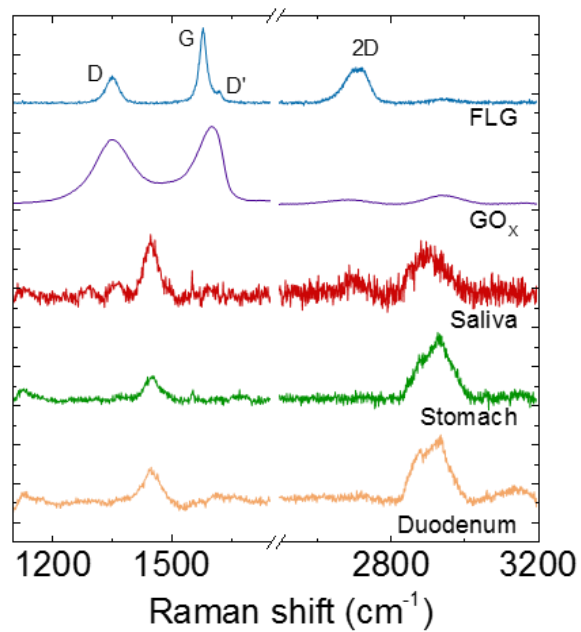

**Figure 3. Representative Raman spectra (@514 nm) of GRMs and digestive juices.** Few-layers graphene and graphene oxide spectra are reported in blue and purple, respectively. Red, green and yellow represent Raman spectra of the digestive juices: saliva, stomach and intestine.

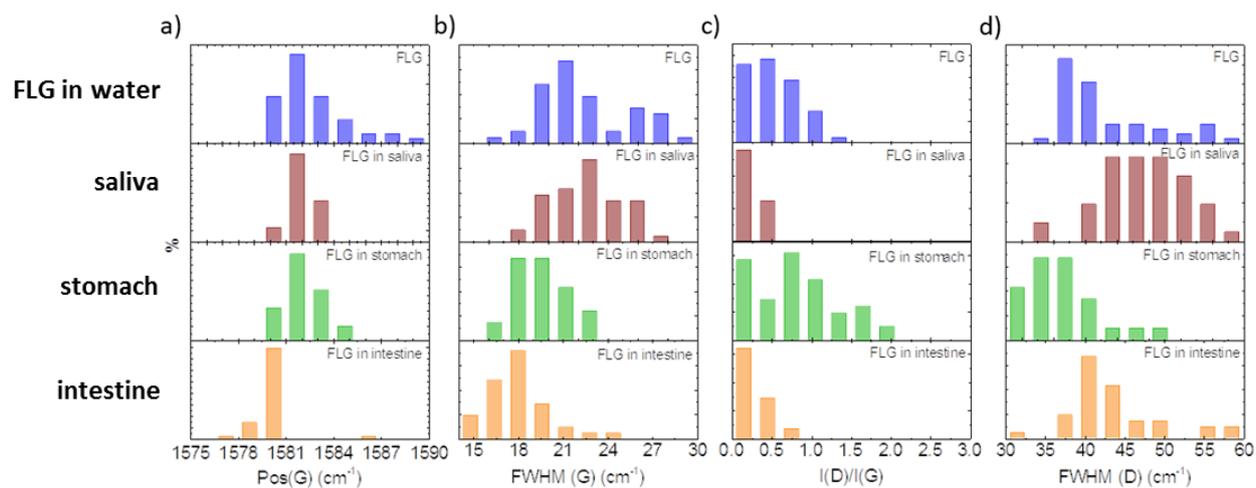

**Figure 4. Raman statistical analysis of FLG dispersions in saliva, stomach and intestine juices.**
a) Pos(G), b) FWHM(G), c) Normalized intensity I(D)/I(G) and d) FWHM(D) of FLG.

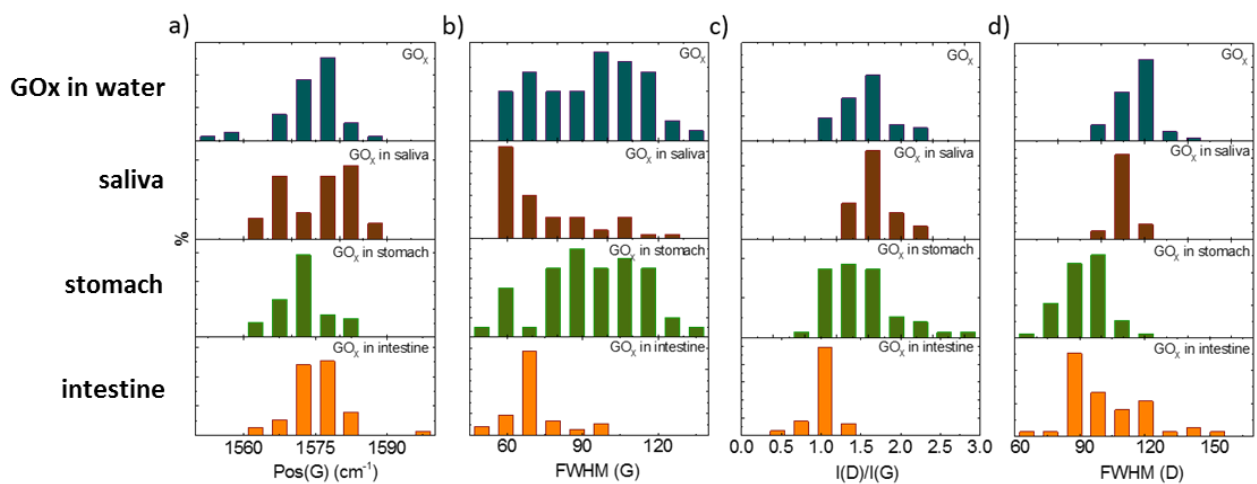

**Figure 5. Raman statistical analysis of GO$_X$ dispersions in saliva, stomach and intestine juices.**
a) Pos(G), b) FWHM(G), c) Normalized intensity I(D)/I(G) and d) FWHM(D) of GO$_X$ flakes.

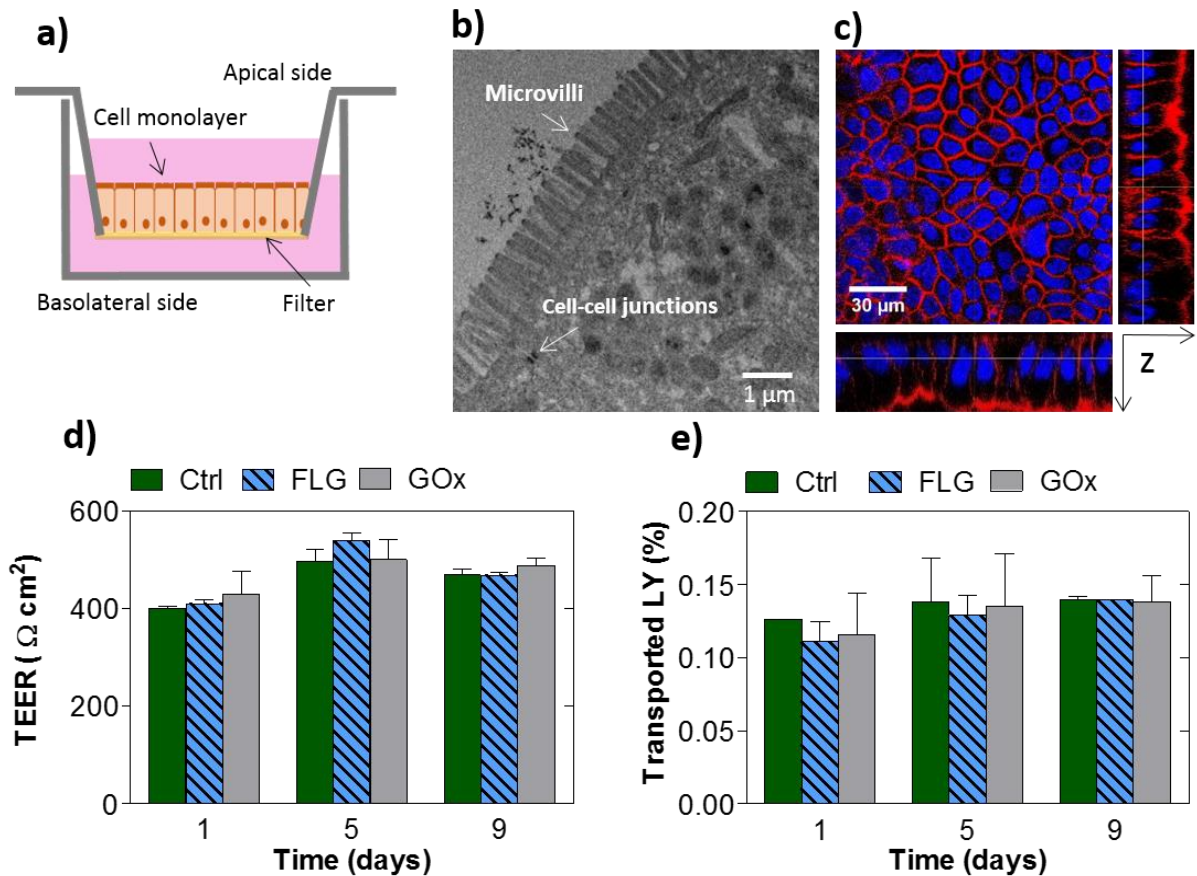

**Figure 6. Characterization of the intestinal epithelium formation and integrity after chronic exposure to digested FLG and GO$_X$.** a) Schematic representation of the intestinal epithelium *in vitro* model; b) TEM micrograph of 21 day-grown intestinal layer showing microvilli and cell-cell junction formation; c) Representative z-sectioning confocal microscopy image of a confluent intestinal layer after 21 days of growth on permeable inserts. Cells are stained with phalloidin (red) and Hoechst 33342 (blue) to highlight actin microfilaments and nuclei. Lateral boxes represent z-stack projections along x-z and y-z axis; d) TEER measurements of non-treated control intestinal epithelial layers (Ctrl) and intestinal epithelial layers after 1, 5 and 9 days of chronic exposure to digested FLG and GOX; e) percentage of transported LY across the intestinal epithelium after 1, 5 and 9 days of chronic exposure to digested FLG and GO$_X$ compared with non-treated control intestinal layers (Ctrl). Data represent the average of three different experiments performed in triplicate and the error bars represent the standard deviation.

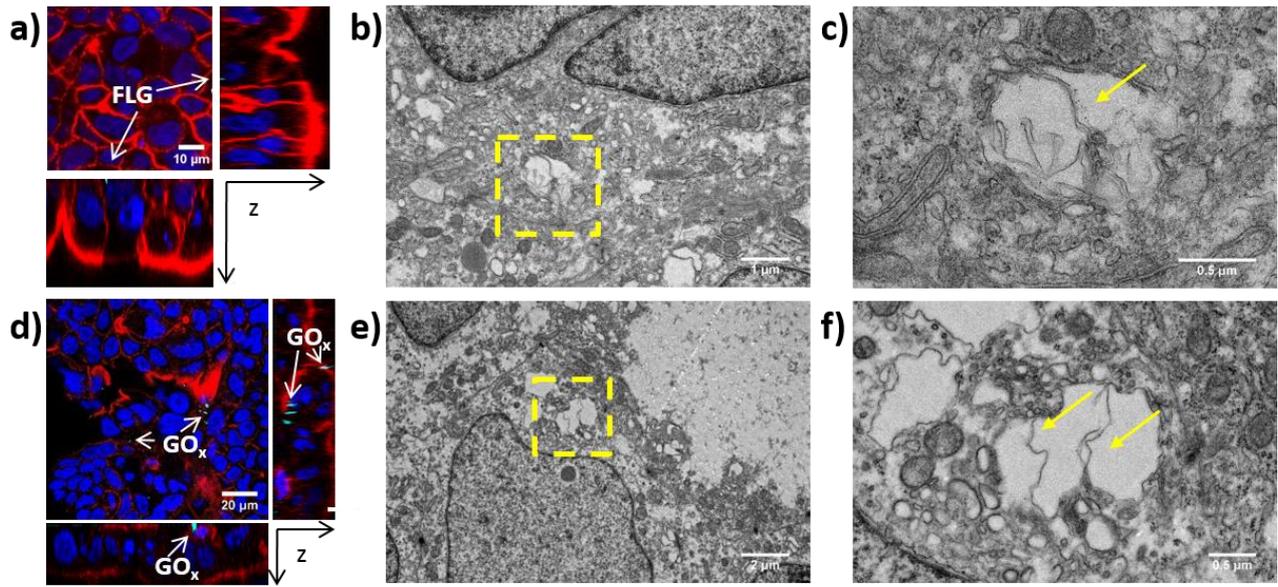

**Figure 7. Uptake and intracellular localization of digested FLG and GO$_X$ nanosheets in Caco-2 intestinal epithelium.** Representative z-sectioning confocal microscopy images of confluent intestinal layers after 9-day chronic incubation with digested FLG (a) and GO$_X$ (d). Cells are stained with phalloidin (red) and Hoechst 33342 (blue) to highlight actin microfilaments and nuclei, respectively, and FLG and GO$_X$ are acquired by reflected light (cyan). Lateral boxes represent z-stack projections along x-z and y-z axis. TEM micrographs of digested FLG (b and c) and GO$_X$ (e and f) nanoflakes internalized in Caco-2 cell barriers after 9-day chronic incubation. c) and f) are the zoomed areas highlighted by the yellow dashed squares in b) and e), respectively. Yellow arrows indicate GRM flakes.

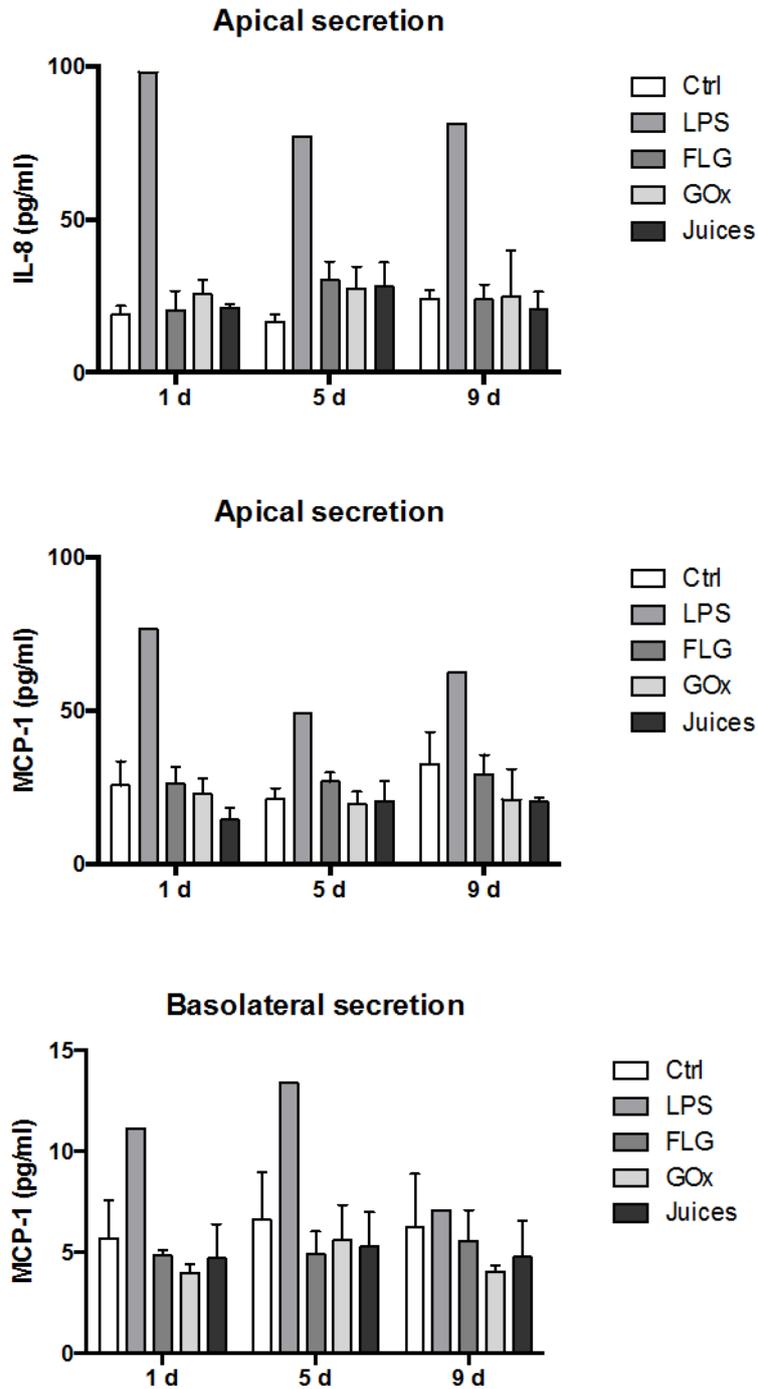

**Figure 8. Inflammatory response of Caco-2 intestinal layer upon chronic exposure to digested FLG and GOx flakes.** Release of IL-8 in apical and basolateral compartments and MCP-1 in basolateral compartment.